\documentstyle[preprint,aps,pra]{revtex}

\begin{document}
\draft
\tighten

\preprint{\parbox[t]{50mm}
{\begin{flushright}
PSI-PR-98-20 \protect\\
ADP-98-55/T323 \protect\\ 
UCY-PHY-98/10  
\end{flushright}}}

\vspace{2cm}

\title{Worldline path integral for the massive Dirac propagator :\protect\\
A four-dimensional approach}

\author{C.~Alexandrou $^{1,2}$, R.~Rosenfelder $^2$ 
and A.~W.~Schreiber $^{2,3}$}

\address{$^1$ Department of Natural Sciences, University of Cyprus, 
CY-1678 Nicosia, Cyprus\\
$^2$ Paul Scherrer Institute, CH-5232 Villigen PSI, Switzerland\\
$^3$ Department of Physics and Mathematical Physics and
           Research Centre for the Subatomic Structure of Matter,\\
           University of Adelaide, Adelaide, S. A. 5005, Australia }

\maketitle

\begin{abstract}
We simplify and generalize an approach proposed by Di Vecchia and Ravndal  
to describe a massive Dirac particle in external vector and scalar fields. 
Two different path integral representations for the propagator are derived
systematically without the usual five-dimensional extension and shown
to be equivalent due to the supersymmetry of the action. They correspond
to a projection on the mass of the particle either continuously or at
the end of the time evolution. It is shown that the supersymmetry
transformations are generated by shifting and scaling the supertimes
and the invariant difference of two supertimes is given for the general 
case. A nonrelativistic reduction of the relativistic propagator leads to
a three-dimensional path integral with the usual Pauli Hamiltonian.
By integrating out the photons we obtain the effective action for quenched 
QED and use it to derive the gauge-transformation properties of the general
Green function of the theory.
\end{abstract}


\section{Introduction}

The problem how to describe spin in a path integral has a long and twisted 
history. This is mostly due to the fact that a path integral is determined
by the classical Lagrangian (or Hamiltonian)
and a classical analogue for the internal spin of a particle 
is not readily available. 
Martin \cite{Mart} apparently first suggested to use anticommuting 
Grassmann
variables for this purpose. This can be made plausible when one recalls 
that the spin operator $ {\bf s} \equiv \hbar {\mbox{\boldmath $\sigma$}}/ 2 $
of an electron fulfills
\begin{equation}
\left \{  s_i, s_j \right \} \equiv s_i \>  s_j
+  s_j \>  s_i \> = \> \frac{\hbar^2}{2} \delta_{i j} \>
\longrightarrow \> 0 \hspace{1cm} i,j = 1,2,3 
\label{spin op}
\end{equation}
in the classical limit. Consequently one can describe a spinning particle
by its bosonic part, the usual trajectory $ x(t) $, and a fermionic
degree of freedom given by a Grassmann valued function $\zeta(t)$
\footnote{It should be noted that there are other approaches, e.g. using
coherent state path integrals \cite{CSPI}, which we will not consider here.}.
Brink {\it et al.} \cite{BDZVH} noted an important supersymmetry between 
the bosonic and fermionic parts of a relativistic massless Dirac particle
and Berezin and Marinov \cite{BeMa}
showed that {\it massive} particles can be described by adding a fifth 
component, $\zeta_5(t)$, to the spin variable. The reason for this 
peculiar addition is that in the rest frame of the particle the spin is 
intrinsically 3-dimensional (cf. Eq.(\ref{spin op})) and a covariant 
4-dimensional description therefore has superfluous degrees of freedom 
which have to be cancelled by the fifth spin variable \cite{Mar}.

There is now a vast literature about spin in path integrals (a partial
list of references is [6-10]) which discusses various aspects of 
this approach. In particular, Fradkin and Gitman \cite{FrGi} have given
a straightforward way of constructing the corresponding relativistic 
propagator.  In addition to the dependence on bosonic and fermionic 
trajectories mentioned above, their formulation has the special feature
that as well as the usual Schwinger proper time a Grassmannian partner to 
it is required.  Representing Dirac particles in a first quantized form in 
the ``worldline formalism'' has become popular for  perturbative 
calculations in QED and QCD [12-14]. These one-loop calculations of the 
effective action are simplified by the fact that only Green functions 
on a circle (with simpler boundary conditions) are needed. More
recently, the method has also been used in order to derive derivative
expansions of the one-loop effective action in $2+1$ and $3+1$ dimensional 
QED~\cite{GusSho}.

Although sufficient for many purposes the Berezin-Marinov introduction
of the fifth spin variable is an awkward one: there is no clear physical
picture associated with it and the corresponding  multiplication of the 
propagator with the Dirac matrix $ \gamma_5$ \cite{FrGi} is very unnatural
in a parity conserving theory.
A four-dimensional approach, which did not get very much attention up to 
now, is that proposed by  Di Vecchia and Ravndal \cite{DiVR,Rav}
in which the unwanted spin degrees of freedom are simply projected out
\footnote{During the course of this work there appeared a publication 
\cite{DIS} in which the fifth spin variable is also eliminated but by a 
different nonlinear technique.}.

It is the purpose of the present paper to develop this latter approach 
further and to show that it has attractive features. In particular, 
in Section~2, we will calculate the propagator for a Dirac particle
in an external vector field and demonstrate that the projection mentioned 
above
can be done in two different ways: either at each time step during the
evolution of the system or at the end. 
We will refer to the former as the ``local''projection method and to 
the latter as the ``global'' projection method.
In Section~3 both procedures are shown to be equivalent due
to the supersymmetry between bosonic and fermionic variables. However, 
the global projection
leads, in general, to simpler expressions without a Grassmann proper 
time. In addition, an inherent coupling between orbital and spin parts 
which is already present for a free particle is removed by the 
global projection method. Section~4 contains the 
nonrelativistic reduction where we start directly from the path integral
representation of the Dirac propagator and show that this reduces to
 the three-dimensional spin path integral of the nonrelativistic theory.
This is to be contrasted with Ref.~\cite{AFP} where  the nonrelativistic
propagator was derived starting  with the 
nonrelativistic Hamiltonian and introducing  three-dimensional Grassmann 
variables instead of obtaining  it
from the relativistic path integral  for the Dirac propagator.
In Section~5 we show that one can also describe a Yukawa 
interaction of the fermion (i.e. the particle in an external scalar 
field) in such a four-dimensional framework. As an application we
derive the effective action in Quantum Electrodynamics in Section~6 and
finally we summarize our results.

Since we aim in making this paper self-contained we include an
Appendix with a derivation of the spin path integral which is 
somewhat different, more explicit and simpler than the one given by 
Fradkin and Gitman. Our conventions follow Bjorken and Drell 
\cite{BjDr} and in general we use
the letters $\alpha-\iota$ and $\xi-\omega$ to denote Grassmann
variables, with some exceptions 
to comply with standard notation found in the literature.

\section{Dirac Propagator in an External Vector Potential}

We are looking for the path integral representation for the propagator 
of a Dirac particle 
\begin{equation}
G(x,y) \> = \>\left < x \> \left | \> \frac{1}{\hat {p \hspace{-5pt}/} - 
g {A \hspace{-6pt}/}(\hat x) - M  + i 0} \>  \right | \> y \right > 
\label{propagator}
\end{equation}
in an external field $ A_{\mu}(x) $ where throughout this paper
quantum mechanical operators are denoted by ``hats'' over the 
corresponding symbols. In the spinless (bosonic) case this can be 
achieved by using Schwinger's proper time representation for the quantum 
mechanical resolvent
\begin{equation}
\frac{1}{E - \hat H + i 0} \> = \>-i \int_0^{\infty} dT \> \exp \left
[ \> i ( E - \hat H + i 0) T \> \right ] \> .
\label{proper time rep}
\end{equation}
However, in the fermionic case we have to make sure that the operator 
$\hat H$
which plays the role of a Hamiltonian for the quantum mechanical system
contains an {\it even} number of Dirac matrices. This is because
in the classical limit (and also in the path integral) only an even, 
commuting object can represent a physical quantity.

Fradkin and Gitman \cite{FrGi} achieved this by multiplying numerator and 
denominator in Eq.~(\ref{propagator}) by $ \gamma_5 $ and extending
the Dirac algebra to five dimensions. However, it is much simpler to use
the representation of Di Vecchia and Ravndal \cite{DiVR,Rav} where we 
write
\begin{equation}
\frac{1}{\hat {\Pi \hspace{-8pt}/} - M + i 0} \> = \> \left ( \> \hat {\Pi \hspace{-8pt}/} + M \> 
\right ) \frac{1}{\hat {\Pi \hspace{-8pt}/}^2 - M^2 + i 0} \> .
\label{DR}
\end{equation}
It is now possible either to exponentiate only the denominator which gives
\begin{equation}
\frac{1}{\hat {\Pi \hspace{-8pt}/} - M + i 0} \> = \> 
 -  \frac{i}{2 \kappa_0}  \int_0^{\infty} dT \> 
\left ( \hat {\Pi \hspace{-8pt}/} + M \right ) \> 
\exp \left (  \> - \frac{i M^2 T}{2 \kappa_0} \> \right )
\> \exp \left (  \> \frac{i}{2 \kappa_0}  \hat {\Pi \hspace{-8pt}/}^2 T \> \right )
\label{DR1}
\end{equation}
or {\it both} numerator and denominator leading to
\begin{equation}
\frac{1}{\hat {\Pi \hspace{-8pt}/} - M + i 0} \> = \>
\int_0^{\infty} dT \int d\chi \>  
\exp \left [  \> - \frac{i}{2 \kappa_0} ( M^2 T + M \chi) \> \right ]
\> \exp \left [  \> \frac{i}{2 \kappa_0} ( \hat {\Pi \hspace{-8pt}/}^2 T + \hat 
{\Pi \hspace{-8pt}/} \chi) \> \right ] \> 
\label{DR2}
\end{equation}
The latter  only holds if $ \> \hat {\Pi \hspace{-8pt}/} \> $
commutes with $ \> \hat {\Pi \hspace{-8pt}/}^2 \> $ which is proved in 
Ref.~\cite {Rav}.
Here
\begin{equation}
\hat \Pi^{\mu} \> = \>\hat p^{\mu}-gA^{\mu}(\hat x) \> ,
\label{kin momentum}
\end{equation}
and the Berezin integrals over Grassmann variables are defined as 
\cite{Ber}
\begin{equation}
\int d\chi  \> = \> 0 \> ,\hspace{1cm} \int d\chi \> \chi  \> = \> 1 . 
\end{equation}
$\kappa_0$ is a parameter which reparametrizes the proper times
 $ T \to \kappa_0 T \> \> , \chi \to \kappa_0 \chi $
without changing the physics and is a remnant of the local
reparametrization invariance of the action. It is thus related to the
``einbein'' \cite{BDZVH}. $\chi$ is either called a (one-dimensional)
``gravitino'' field or, more appropriately in the present context,
as the supersymmetric partner of the proper time, the ``supertime''.

The Di Vecchia-Ravndal representation has several advantages compared to 
the standard Berezin-Marinov form \cite{BeMa} for the description of a 
{\it massive} spinning particle: no five-dimensional extension and 
multiplication with $\gamma_5$ are necessary and, as we will see in 
Section~3, the supersymmetric transformations  are much simpler
and more transparent.  
It can be considered as
the result obtained by integrating out the fifth spin variable. 
A certain disadvantage is that not all exponents in Eq.~(\ref{DR2}) are 
Grassmann even. The odd term 
\begin{equation}
\exp \left ( -\frac{i M}{2 \kappa_0} \chi \right )
\end{equation}
is to be considered as part of an operator which projects out 
$\hat {\Pi \hspace{-8pt}/} = M$ 
\cite{Rav} and not as part of the evolution operator. In Eq.~(\ref{DR1}) 
this projection is done at the end (``global'') 
whereas in Eq.~(\ref{DR2}) it is done 
at each time step during the evolution (``local''). 

It is essential that in both procedures the ``Hamiltonian'' which governs
the proper-time evolution is even. In the global projection method it is 
given by
\begin{equation}
 {\cal H}(\hat \Pi,\hat x,\gamma) \> = \>  - \frac{1}{2 \kappa_0}
 \hat {\Pi \hspace{-8pt}/}^2 \> = \>-\frac{\hat \Pi^2}{2 \kappa_0} + 
\frac{i}{4 \kappa_0} \, g \, F_{\mu\nu}(\hat x) \, \gamma^{\mu} 
\gamma^{\nu} \, , \> \> \> \left ( \> F_{\mu\nu} \equiv \partial_{\mu} A_{\nu} 
- \partial_{\nu} A_{\mu} \> \right  )
\label{Hamiltonian}
\end{equation}
whereas for the local projection method it reads
\begin{equation}
 {\cal H'}(\hat \Pi,\hat x,\gamma) \> = \>{\cal H}(\hat \Pi,\hat x,\gamma)
 - \frac{1}{2 \kappa_0 T} \hat \Pi_{\mu} \gamma^{\mu} \chi \> . 
\label{Hamiltonian'}
\end{equation}
In both cases the parameter $\kappa_0$ can be interpreted as the ``mass''
of the quantum-mechanical particle.

\subsection{Global projection}

We will first consider the projection after the end of the evolution, i.e.
\begin{eqnarray}
G(x,y) &=& - \frac{i}{2 \kappa_0} \int d^4 z \left < x \left | 
\hat {p \hspace{-5pt}/} - g {A \hspace{-6pt}/} (\hat x) + M \right | z \right > \, 
\int_0^{\infty} dT  \> 
\exp \left (  \, - \frac{i M^2 T}{2 \kappa_0} \, \right )
\, \left < z \left | \exp \left (  \, - i \hat {\cal H} \, T \, \right ) 
\right | y \right > \nonumber \\
&=& - \frac{i}{2 \kappa_0} \Bigl ( \, i \partial \hspace{-6pt}/_x - g {A \hspace{-6pt}/}(x) + M \, 
\Bigr ) \int_0^{\infty} dT  \> \exp \left (  \> - 
\frac{i M^2 T}{2 \kappa_0} \> \right ) \> \left < x \left | \exp 
\left (  \> - i \hat {\cal H} \> T \> \right ) \right | y  \right > \>.
\end{eqnarray}
The remaining proper-time evolution operator
can be written in path integral form following 
Fradkin and Gitman \cite{FrGi} but staying within a four-dimensional
framework. In Appendix A we show that
\begin{eqnarray}
\left < x \left | \exp \left ( - i \hat {\cal H} \> T \right ) \right |
y \right > &=& \exp\left ( \gamma
\cdot \frac{\partial}{\partial \Gamma} \right )
\int_{x(0)=y}^{x(T)=x} {\cal D}x \> {\cal D}p  \> {\cal D}\xi \> 
N^{\, \rm spin} \nonumber \\
&&  \cdot \> \exp \left \{ i \int_0^T \> dt
\biggl [ i \xi \cdot \dot \xi \> - p \cdot \dot x - 
{\cal H}(\> \Pi,x,2 \xi + \Gamma)\> ) \> \biggr ]  \> \right \}
_{\Gamma = 0} \> ,
\label{phase space PI1}
\end{eqnarray}
where $N^{\, \rm spin}$ is a normalization factor for the 
four-dimensional spin path integral as given in Eq.~(\ref{N spin}) and 
we use  antiperiodic boundary conditions for the spin  variable $\xi(t)$ 
\begin{equation}
\xi_{\mu}(0) + \xi_{\mu}(T) \> = \>0 \>.
\end{equation}
Eq.~(\ref{phase space PI1}) can be further simplified by shifting to the 
new spin variables
\begin{equation}
\zeta_{\mu}(t) \> = \>\frac{1}{2 } \Gamma_{\mu} + \xi_{\mu}(t) 
\label{def zeta}
\end{equation}
so that the boundary condition becomes
\begin{equation}
\zeta_{\mu}(0) + \zeta_{\mu}(T) \> = \>\Gamma_{\mu} \> .
\label{bound cond zeta}
\end{equation}
This introduces an additional boundary term: 
$ \> -  \frac{1}{2} \Gamma \cdot \left [ \, \zeta(T) - \zeta(0) \, \right ] 
= \zeta(T) \cdot \zeta(0) \> $.
After shifting to the momentum (\ref{kin momentum}) as 
integration variable we  obtain
\begin{eqnarray}
G(x,y)  &=&  - \frac{i}{2 \kappa_0} \Bigl ( i \partial \hspace{-6pt}/_x 
- g {A \hspace{-6pt}/}(x)  + M  \Bigr )  \exp \left ( \gamma \cdot \frac{\partial}
{\partial \Gamma}\right ) \int\limits_0^{\infty} dT  
\exp \left ( - \frac{i}{2 \kappa_0} M^2 T  \right )
N^{\, \rm spin} \int{\cal D}x  {\cal D}\Pi   {\cal D}\zeta 
\nonumber \\
&& \cdot  \exp \left \{  \> \zeta(0) \cdot \zeta(T) \> 
 + i \int_0^T dt \> \biggl [ \> i \zeta 
\cdot \dot \zeta - (\Pi + g A(x)) \cdot \dot x - {\cal H}(\> \Pi,x, 
2 \zeta \> ) \> \biggr ]  \> \right \}
_{\Gamma = 0}  .
\label{phase space PI2}
\end{eqnarray}
As usual we can perform the functional $\Pi$-integration since the 
Hamiltonian is at most quadratic in the kinematical  momentum. We then 
obtain the final expression
\begin{eqnarray}
 G(x,y) &=& - \frac{i}{2 \kappa_0} \Bigl( \, i \partial \hspace{-6pt}/_x - g {A \hspace{-6pt}/}(x) 
+ M \, \Bigr )  \exp \left ( \gamma \cdot \frac{\partial}
{\partial \Gamma} \right ) \int_0^{\infty} dT \> N(T) \exp \left (  
\> - \frac{i}{2 \kappa_0}  M^2 T \> \right ) \>  \nonumber \\
 && \hspace{5cm} \cdot \> \int{\cal D}x {\cal D}\zeta \>  
\exp \Bigl \{ \> i S [x,\zeta]  \> \Bigr \}_{\Gamma = 0}
\label{G propagator}
\end{eqnarray}
where
\begin{equation}
N(T) \> = \>\left [ \int {\cal D}\zeta \> \exp \left ( \zeta(0) \cdot 
\zeta(T) - \int_0^T dt \> \zeta \cdot  \dot \zeta \right ) \right ]^{-1}
 \> \cdot \> \int {\cal D}\Pi ~\exp \left ( i \int_0^T dt \>
\frac{\Pi^2}{2 \kappa_0} \right ) 
\end{equation}
provides the proper normalization and
\begin{eqnarray}
S[x,\zeta]  & \equiv &  \int_0^T dt \> L(x,\dot x,\zeta,\dot \zeta) 
\> - \> i \zeta(0) \cdot \zeta(T) \nonumber \\
L(x,\dot x,\zeta,\dot \zeta) &=& -\frac{\kappa_0}{2} \dot{x}^2 +  
i \zeta \cdot \dot \zeta - g \dot{x} \cdot A(x) - \frac{ig}{\kappa_0} 
F_{\mu\nu} (x) \, \zeta^{\mu}\zeta^{\nu} 
\label{action in x-space}
\end{eqnarray}
are the action and the Lagrangian, respectively.  The first two terms in
Eq.~(\ref{action in x-space}) correspond, respectively, to contributions 
from the orbital and
spin degrees of freedom to the kinetic energy, while the last two terms 
are the contributions of the photon field
coupling to both the electron's convection current and its
spin current.
The canonically conjugate 
momenta are given by
\begin{equation}
p_{\mu} \> = \>\frac{\partial L}{\partial \dot x^{\mu}} \> = \>
- \kappa_0 \dot x_{\mu} - g A_{\mu}(x) 
\end{equation}
and
\begin{equation}
\eta_{\mu} \> = \> \frac{\partial L}{\partial \dot \zeta^{\mu}} \> = \> 
- i \zeta_{\mu}
\end{equation}
so that the canonical Hamiltonian becomes
\begin{equation}
H \> = \>\sum_{q_i = x,\zeta} \dot q_i \frac{\partial L}{\partial \dot q_i} 
- L \> = \>- \frac{\kappa_0}{2} \dot x^2 + \frac{ig}{\kappa_0} F_{\mu \nu} 
\zeta^{\mu} \zeta^{\nu} 
\label{canon Hamiltonian}
\end{equation}
when expressed in terms of (generalized) coordinates and velocities. 
In terms of coordinates and momenta we have the relation $ H = 
{\cal H}|_{p \to -p}$.
This is a consequence of our metric which gives $ \> \exp(-i p \cdot x) 
\> $ as plane wave and therefore leads to the form 
$ \> \int dt \> [ - p \cdot \dot x - {\cal H} ] \> $ for the action 
in the phase space path integral (\ref{phase space PI1}).

The free Dirac propagator in momentum space is readily obtained from 
(\ref{G propagator})
 since orbital and spin variables decouple. The 
$\zeta$-path integral cancels against the normalization factor 
$ N^{\, \rm spin} $ and the $x$-path integral gives just the usual free
bosonic evolution kernel. Thus
\begin{eqnarray}
G^{(0)}(p) &=& \int d^4x \> e^{ i p \cdot x} \left (- \frac{1}{2 \kappa_0}
\right ) \left ( i \partial \hspace{-6pt}/_x  + M  \right ) \int_0^{\infty} dT \>
\exp \left ( - \frac{i}{2 \kappa_0} M^2 T   \right) \nonumber \\
&& \cdot \int \frac{d^4k}{(2 \pi)^4} 
\> e^{-i k \cdot x} \exp \left (  \frac{ i}{2 \kappa_0} k^2T  \right ) 
 \> = \> \frac{ {p \hspace{-5pt}/} + M }{ p^2 - M^2 + i 0} \> = \>
\frac{1}{{p \hspace{-5pt}/} - M + i0} \> ,
\label{free propagator}
\end{eqnarray}
independent of the reparametrization parameter $\kappa_0$.

\subsection{Local projection}

The local projection method follows along the same lines with two 
differences: first we have an additional integration over the 
supertime $\chi$ and second there is an extra term in the action due 
to the additional term in Eq.~(\ref{Hamiltonian'}). Thus
\begin{eqnarray}
 G\>'(x,y) &=& \exp \left ( \gamma \cdot \frac{\partial}{\partial \Gamma} 
\right ) \int_0^{\infty} dT \> N(T) \int d\chi \> 
\exp \left [  \> - \frac{i}{2 \kappa_0}  ( M^2 T + M \chi ) \> \right ] 
\nonumber \\
&& \hspace{4cm} \cdot \int{\cal D}x \> {\cal D}\zeta 
 \>  \exp \Bigl \{ \> i S \> ' [x,\zeta]  \> \Bigr \}_{\Gamma = 0}
\label{G propagator local}
\end{eqnarray}
with 
\begin{eqnarray}
S\>' [x,\zeta] & \equiv & \int_0^T dt \> L'(x,\dot x, \zeta,\dot \zeta; 
\chi) \> - \> i \zeta(0) \cdot \zeta(T) \nonumber \\
L\>'(x,\dot x, \zeta,\dot \zeta; \chi) &=& L (x,\dot x, \zeta,\dot \zeta)
\> + \> \frac{1}{T} \> \dot x^{\mu} \zeta_{\mu} \> \chi 
\label{action' in x-space}\\
&=& - \frac{\kappa_0}{2} \dot{x}^2 +
i \zeta \cdot \dot \zeta + \> \frac{1}{T} \> \dot x \cdot \zeta \chi
- g \dot{x} \cdot A(x) - \frac{ig}{\kappa_0}
F_{\mu\nu} (x) \, \zeta^{\mu}\zeta^{\nu} \nonumber \> .
\end{eqnarray}
Note that there is now a coupling between orbital movement and spin, even 
for the free particle. This is the same mechanism which at high energy 
aligns the spin of a Dirac particle along (or opposite) to the momentum
whereas a non-relativistic particle with spin is unaffected.

Since the spin degrees of freedom appear at most quadratically
it is also possible to integrate them out completely and reduce the path
integral to a bosonic one modified by a ``spin factor''
\cite{Pol,spin factor}. The price to be paid is that
this spin factor is highly nonlinear in
the external fields. This prevents an analytic integration over the
boson fields to obtain an effective interaction 
for the fermion only, as is done in Section~6.

\section{Bosonic and Fermionic Transformations}

We next discuss the transformation properties of the Lagrange function 
in the local formulation \cite{BDZVH,FaMa}. 
The corresponding ones for the global
formulation can be obtained by setting $ \> \chi = 0 \> $. 
There are two kinds of transformations which leave the
 Lagrange function $L'$ in Eq.~(\ref{action' in x-space}) 
invariant (up to a total derivative):
\begin{itemize}
\item[(i)] bosonic transformations (reparametrizations)
\begin{eqnarray}
\delta x^{\mu} &=& b(t) \> \dot x^{\mu} \nonumber \\
\delta \zeta^{\mu} &=& b(t) \> \dot \zeta^{\mu} 
\label{bosonic transf} \\
\delta \kappa_0 &=& - \kappa_0^2 \frac{d}{dt} \left ( \> 
\frac{b(t)}{\kappa_0} \> \right )\nonumber \\
\Rightarrow \delta L' &=& \> \frac{d}{dt} \left [ \> b(t) \> L' 
\> \right ] \;\;\;,
\label{dL0 boson}
\end{eqnarray}
where $ b(t) $ is the infinitesimal parameter of the 
transformation which, in principle, could have an arbitrary 
time-dependence (local transformations).
However, for quantization the reparametrization ``gauge'' has to be 
fixed \cite{BeMa,BVH} which in our case, by construction, was taken to 
be $\kappa_0 =$ constant.
This means that we only can allow $\delta \kappa_0 =$ constant, or
\begin{equation}
b(t) \> = \>b_0 +  b_1 \> t \> .
\label{allowed b(t)}
\end{equation}
Note that $b_1 = 0$ corresponds to proper time {\it translations}, e.g.
$ \> \delta x^{\mu} = x^{\mu}(t+b_0) - x^{\mu}(t) = b_0 \> 
\dot x^{\mu} \> + ...\> $, and
$b_0 = 0$ to proper time {\it scalings}, e.g.
$ \> \delta x^{\mu} = x^{\mu}(t+b_1 t) - x^{\mu}(t) = b_1 t \> 
\dot x^{\mu} \> + ... \> $. 

\item[(ii)] fermionic (supersymmetric) transformations 
\begin{eqnarray}
\delta x^{\mu} &=& i \alpha(t) \> \zeta^{\mu} \nonumber \\
\delta \zeta^{\mu} &=&  \frac{\kappa_0}{2} \alpha(t) \> \left [ \>  
\dot x^{\mu} - \frac{1}{\kappa_0 T} \zeta ^{\mu}\chi \> \right ] \nonumber \\ 
\delta \kappa_0 &=& \frac{\kappa_0}{T} \alpha(t) \chi \nonumber \\
\delta \chi  &=& - i \kappa_0 T \dot \alpha(t)
\label{fermionic transf}  \\
\Rightarrow \delta L' &=&  i \> \frac{d}{dt} \left [ \> \alpha(t)
\left ( - \frac{\kappa_0}{2} \dot x \cdot \zeta \> + \> g \> A \cdot 
\zeta \> \right ) \> \right ] \;\;\;,
\label{dL0 fermion}
\end{eqnarray}
where $ \> \alpha(t) \> $ is the infinitesimal Grassmannian 
parameter of the transformation. Again, since $ \> \kappa_0$ and  
$\chi \> $ are by construction time-independent, one can only allow 
transformations with $ \> \alpha(t) = \> $ constant. 
For $ \chi = 0 $ Ravndal \cite{Rav} has shown that 
similar to the bosonic case it is also possible to generate the 
fermionic transformations by a shift in the proper time if  
a Grassmannian partner of the proper time $ t $ is added. This allows
for a concise supersymmetric formulation of the action. 
In this Section we will show the generalization of Ravndal's 
transformations to the case with spin-orbit coupling, which includes  
a special scaling of the ``supertime'' in addition to a shift.

\end{itemize}
Since the change of the full Lagrange function is a total
derivative, Noether's theorem~\cite{ItZu} allows us to define quantities 
which  are conserved classically.
For the bosonic transformation with $b_1 = 0$ (i.e. proper time 
translations) we have $\delta \kappa_0 = 0$ and 
Eq.~(\ref{dL0 boson})  therefore leads to the conservation of the 
canonical Hamiltonian (\ref{canon Hamiltonian}).
It can be shown that proper time scalings ($b_0 = 0$) 
lead to the same result \footnote{In this case the reparametrization 
parameter $\kappa_0$, which is also changed, is not a dynamical variable
for which the equations of motion can be used. Consequently the change of 
the corresponding Noether charge $Q = t H $ with time is proportional to 
$ \kappa_0 \partial L /\partial \kappa_0 = H$, which gives no new 
information.}. 
For the fermionic transformations we find from Eq.~(\ref{dL0 fermion}) 
that the projection of the spin variable on the 
kinematical momentum $-\kappa_0 \dot x $
\begin{equation}
Q \> = \> - \kappa_0 \dot x_{\mu} \zeta^{\mu}
\label{def Q}
\end{equation}
is conserved classically without the spin-orbit term \cite{Rav}.

Quantum mechanically the Noether charges either become conserved 
operators or, in the functional formalism, their conservation implies
that certain averages, i.e. Green functions
with the Noether charges as insertions, stay time-independent.
For quantum mechanical averages we will use the following notation 
\begin{equation}
\left < {\cal O} \right >_S \> \equiv \> 
\exp \left ( \gamma \cdot \frac{\partial}{\partial \Gamma} \right )
\int {\cal D}x \> {\cal D} \zeta \> {\cal O}(x,\zeta) \> 
e^{ i S[x, \zeta]} \> \biggr|_{\Gamma=0} \> .
\label{def average}
\end{equation}
To be specific, we consider the fermionic transformations with 
$ \> \chi = 0 \> $ because their
Noether charge (\ref{def Q}) does not dependent (explicitly) on the 
interaction and we  make a {\it local}, time-dependent 
transformation~\cite{PeSch} 
\begin{equation}
x(t) \> = \>x'(t) + i \> \alpha(t) \> \zeta (t) \> , \hspace{1.5cm} 
\zeta(t) \> = \>\zeta'(t)
+ \frac{\kappa_0}{2} \> \alpha(t)\> \dot x(t)
\end{equation}
in the path integral. We assume that $\alpha(0) = \alpha(T) = 0$
so that we do not have to consider boundary contributions. The 
Jacobian for this transformation is $ 1 + {\cal O} (\alpha^2)$. Since the 
path integral does not change its value we obtain (omitting the primes)
\begin{equation}
0 \> = \> \left < \> i \int_0^T \> dt \> \delta L \> \right >_S 
\end{equation}
where
\begin{equation}
\delta L \> = \> i \> \alpha(t) \frac{d}{dt} \left ( \> - 
\frac{\kappa_0}{2} \dot x \cdot \zeta - g A \cdot \zeta \> \right ) \> 
\>  + i \> \dot \alpha(t) \left ( \> - \frac{3 \kappa_0}{2} \dot x \cdot 
\zeta - g A \cdot \zeta \> \right ) \> .
\end{equation}
The first term is what we obtain for a global, time-independent 
transformation in Eq.~(\ref{dL0 fermion}).
Performing an integration by parts in the second term (no boundary terms)
the result is then
\begin{equation}
0 \> = \>\left < \>
\int_0^T  dt \> \alpha(t)
\left [ \> - \kappa_0 \frac{d}{dt} \left (  \dot x \cdot \zeta \right ) \> 
\right ] \> \right >_S
\end{equation}
or since $ \> \alpha(t) \> $ is arbitrary
\begin{equation}
\frac{d}{dt} \, \Bigl < \> - \kappa_0 \dot x(t) \cdot \zeta(t) \> \Bigr >_S  
\label{conserved Q average} \> = \>0 
\end{equation}
for all times.

\subsection{Supersymmetric formulation}

It is convenient to write the Lagrange function for a relativistic
spinning particle
in explicit supersymmetric form by combining
orbital and spin degrees of freedom into a ``superfield'' \cite{BDZVH}
or ``superposition'' \cite{Rav}
\begin{equation}
X^{\mu}(t,\theta) \> = \>x^{\mu}(t) + \> a \>  \theta \> \zeta^{\mu}(t) \> .
\label{superfield}
\end{equation}
Here $ \theta $ is an additional time-independent Grassmann variable 
which acts as a superpartner of the proper time $t$ 
and $ a $ a suitably chosen constant. If, in addition, a 
``superderivative'' is defined as 
\begin{equation}
D \> = \>\frac{\partial}{\partial \theta} - \theta 
\frac{\partial}{\partial t}
\label{superderivative}
\end{equation}
then 
\begin{equation}
L_0 \> = \> \int d\theta \> \left ( - \frac{\kappa_0}{2} \right ) \> 
D X_{\mu} D^2 X^{\mu} \> = \> - \frac{\kappa_0}{2} \dot x^2
+ i \zeta \cdot \dot \zeta 
\label{free susy L}
\end{equation}
generates all terms in the free Lagrangian of the spinning 
particle provided the constant $ a $ is chosen as
\begin{equation}
a \> = \>i \sqrt{\frac{2 i}{\kappa_0}} \> .
\end{equation}
Note that in this compact form only first-order derivatives appear since
$ D^2 = - \partial/\partial t$. 

\noindent
In the local projection
approach one needs an additional $\chi$-dependent factor \cite{BDZVH,MNSS1} 
\begin{equation}
e(\theta \chi) \> = \>1 + \frac{a}{i T} \theta \chi
\label{def e-factor}
\end{equation}
in the integrand of Eq.~(\ref{free susy L}) to account for
the explicit spin-orbit coupling. Thus the corresponding free action is
\begin{equation}
S_0\> '[X] \> = \>\int_0^T dt \int d\theta \> e(\theta \chi)
 \> \left ( - \frac{\kappa_0}{2} \right ) \> D X \cdot D^2 X
\> = \>\int_0^T dt \> \left [ \>  - \frac{\kappa_0}{2} \dot x^2
+ i \zeta \cdot \dot \zeta + \frac{1}{T} \zeta \cdot \dot x \chi \> 
\right ] \>.
\label{susy S0'}
\end{equation}
\noindent
The interaction of the Dirac particle with an electromagnetic 
field takes the equally simple form
\begin{equation}
L_{\rm e.m.} \> = \>g \int d\theta \> D X_{\mu} A^{\mu} (X)
\label{susy em inter}
\end{equation}
which is easily proved by expanding the ``superposition'' $X$ and 
performing the Berezin integration. Eq.~(\ref{susy em inter}) thus 
contains {\it both} the convection current and the spin current 
interaction. 

For $ \> \chi = 0 \> $
Ravndal and Di~Vecchia \cite{DiVR,Rav} have given a simple way of 
generating both the bosonic (with $b_1=0$)  as well as the fermionic 
transformations by a shift in the proper times $t$ and $\theta$:
\begin{eqnarray}
t  &\rightarrow& t \> + \> b_0 \> + \>   \epsilon \, \theta \nonumber \\
\theta &\rightarrow& \theta \> + \> \epsilon \;\;\;,
\label{susy transf chi = 0}
\end{eqnarray}
where $\epsilon$ and $b_0$ are constants which may be zero.
Indeed, the superfield changes into
\begin{eqnarray}
X(t,\theta) \> \rightarrow \>  X \> ' (t,\theta) &=& x(t+b_0+\epsilon 
\theta) + a (\theta + \epsilon) \zeta(t+b_0+\epsilon \theta) \nonumber \\
&=& x + b_0 \dot x + a \epsilon \zeta + a \theta \left [ \> 
\zeta  + b_0 \dot \zeta - \frac{\epsilon}{a} \dot x \> 
\right ] \> + ... \>  ,
\end{eqnarray} 
and if we set 
\begin{equation}
\epsilon \> = \>\frac{i}{a} \alpha
\label{eps alpha}
\end{equation}
we obtain
both transformations (\ref{bosonic transf}) and
(\ref{fermionic transf}) for the individual 
components of the superfield in the special case $ \> \chi = 0 \> $.
This is not only more transparent but also
treats bosonic and fermionic transformations on an equal footing.
The equations of motion and the conserved quantities can also be 
formulated compactly in this formalism. 

\noindent
We can generalize
the transformations (\ref{susy transf chi = 0}) to the case 
$ \> \chi \neq 0 \> $ by observing that any change in $ \> t, \theta \> $
leaves $ \> \kappa_0, \chi \> $ unchanged, since these quantities are by
construction time-independent. This means that necessarily
\begin{eqnarray}
\delta \kappa_0 &=& 0 \\
\delta \chi &=& 0 \> .
\label{delta kappa0,chi = 0}
\end{eqnarray}
While the latter condition is fulfilled by a constant parameter 
$ \> \alpha \> $ in the fermionic
transformation (see Eq.~(\ref{fermionic transf})) the former one requires 
that the bosonic scaling parameter $ \> b_1 \> $ is not arbitrary but 
given by
\begin{equation}
b_1 \> = \>\frac{1}{T} \> \alpha \chi \> .
\end{equation}
Using Eq.~(\ref{eps alpha}) we then find that
\begin{eqnarray}
t' &=& \left ( 1 + \frac{a}{i T} \epsilon \chi \right ) t + b_0 + 
\epsilon \, \theta \nonumber \\
\theta' &=& \left ( 1 + \frac{a}{2 i T} \epsilon \chi \right ) \theta + 
\epsilon  
\label{susy transf with chi}
\end{eqnarray}
generate the $\chi$-dependent supersymmetric transformations with
$ \> \delta \kappa_0 = \delta \chi = 0 \> $. Although this constitutes
a scaling of the bosonic time $ t $ by a factor
\begin{equation}
\ell \> = \>\left ( 1 + \frac{a}{i T} \epsilon \chi \right )
\label{def l}
\end{equation}
the fermionic time $ \theta $ is only scaled by $ \> \sqrt{\ell} \> $.
Consequently $ \> D \> $ scales by $ \> 1/\sqrt{\ell} \> $. Since
the spin-orbit factor (\ref{def e-factor})
scales again with $ \> \ell \> $ and the Berezin integral over $ \theta $
transforms inversely compared to a bosonic one
the free action is easily found to be invariant under scaling.

\noindent
We also note that for two times $ \> t_1, t_2, \theta_1, \theta_2 \> $
the combination
\begin{equation}
T_{12} \> \equiv \> \frac{t_1 - t_2}{\sqrt{ e(\theta_1 \chi)\> 
e(\theta_2 \chi)}} + \theta_1 \theta_2
\label{invariant time diff}
\end{equation}
is invariant under the shift and scaling (\ref{susy transf with chi}) of 
proper times. 
This is the generalization of a result which is well known for
$ \> \chi = 0 \> $ \cite{Pol} and is
important for extensions of the polaron variational approach to QED
\cite{ARS}.

\subsection{Equivalence of local and global projection}

We are now able to prove the equivalence between the local projection 
method and the global one. We give here a somewhat different and more 
explicit derivation than the one sketched in 
Ref.~\cite{RSchSch}.
We start from the local formulation and perform the $\chi$ - integration.
This gives
\begin{eqnarray}
G \> ' (x,y) &=& 
\exp \left ( \gamma \cdot \frac{\partial}{\partial \Gamma} \right )
\> \int_0^{\infty} dT \> N(T) \exp \left ( - \frac{i}{2 \kappa_0} M^2 T 
\right) \nonumber \\
&& \cdot \int_{x(0)=y}^{x(T)=x} {\cal D}x 
\int_{\zeta(0)+\zeta(T)=\Gamma} {\cal D} \zeta \> e^{i S} \> 
\left [ \> - \frac{i}{2 \kappa_0} M
 - \frac{i}{T} \int_0^T dt \> \dot x \cdot \zeta \right ]_{\Gamma=0}\>. 
\end{eqnarray}
As we have seen in Eq.~(\ref{conserved Q average})  the supersymmetry of 
the action $S$ leads to the result that the expectation
value of $ \dot x \cdot \zeta $ is time-independent and thus can be 
evaluated at any time $t$, in particular at $t = T$. We then can perform 
the $t$-integral and obtain
\begin{equation}
G \> ' (x,y) \> = \>\> \int_0^{\infty} dT \> N(T) \exp \left ( - 
\frac{i}{2 \kappa_0} M^2 T \right) \> \left <  - \frac{i}{2 \kappa_0} M
 - i\> \dot x(T) \cdot \zeta (T) \right >_S 
\label{G' with susy}
\end{equation}
where the average with respect to the action $S$ is defined in 
Eq.~(\ref{def average}).
For the calculation of the last average we use the well-known 
fact (see e.g. Ref.~\cite{ChLi}) that the expectation value of
time-ordered products of Heisenberg operators
$ \> \hat {\cal O}_H(t) = \exp( i \hat {\cal H} t ) \, \hat {\cal O} \> 
\exp(- i \hat {\cal H} t) \> $ is given by the insertion of 
${\cal O}(t)$ in the corresponding path integral. Thus
\begin{equation}
< x,T |\> {\cal T} \left  [ \hat x_H(t_1) \cdot 
\frac{ \hat \gamma_H(t_2)}{2} \right ] \> 
| y, 0 >  \equiv \> < x | \> e^{- i \hat {\cal H} T}
\> {\cal T} \left [ \hat x_H(t_1) \cdot \frac{ \hat \gamma_H(t_2)}{2}  
\right ]\> | y > \> = \>\left < \> x(t_1) \cdot \zeta(t_2) \> \right >_S
\end{equation}
since Eq.~(\ref{phase space PI2}) tells us that the (Weyl ordered) 
$\gamma$ matrices are to be replaced by $2 \zeta$. Differentiating 
with respect to $t_1$ (the equal time contribution vanishes) and 
putting $t_1 = t_2 = T$ we obtain
\begin{equation}
< \dot x(T) \cdot \zeta(T) >_S \> = \>
< x | \> i \left [ \hat {\cal H}, \hat x_{\mu} \right ] 
\frac{ \gamma^{\mu}}{2} \> e^{-i \hat {\cal H} T} \> | y > \> .
\end{equation}
Evaluating the commutator with the help of Eq.~(\ref{Hamiltonian}) and 
the canonical commutation relations we find
\begin{equation}
< \dot x(T) \cdot \zeta(T) >_S   \> = \>- \frac{1}{2 \kappa_0} < x | \> 
\hat {\Pi \hspace{-8pt}/} \> e^{-i \hat {\cal H} T} \> | y > \> = \>\frac{i}{2 \kappa_0}
\Bigl [ \partial \hspace{-6pt}/_x + i g {A \hspace{-6pt}/}(x) \Bigr ] \, < x | \> 
e^{-i \hat {\cal H} T} \> | y > \>
\end{equation}
which, inserted into Eq.~(\ref{G' with susy}),
gives exactly the same result for the propagator as the global projection
method, i.e.
\begin{equation}
G \> '(x,y) \> = \>G(x,y) \> .
\end{equation}

\section{Nonrelativistic Limit}

If the mass $M$ of the fermion becomes large the integral
over the proper time $T$ is dominated by the stationary points
of its integrand which approximately occur at $ \Pi_0^2 = M^2 $. 
Therefore we make the {\it Ansatz}
\begin{equation}
\Pi_0 \> = \>s \, M + E \> ,\hspace{0.5cm} s \> = \>\pm 1 \> .
\end{equation}
For the nonrelativistic limit it is very convenient and natural to take
\begin{equation}
\kappa_0 \> = \>M
\end{equation}
and to assume
\begin{equation}
E \> = \>{\cal O} \left ( \frac{1}{M} \right ) \>.
\end{equation}
In this Section we will write $ {\cal D}^d x , \>  {\cal D}^d p$  and
$N_d^{\, \rm spin} $ with $ d = 3,4 $ to stress the different 
dimensionality of relativistic and nonrelativistic path integrals.
In the global projection method we then obtain
from Eq.~(\ref{phase space PI2}) 
\begin{eqnarray}
G(x,y) &\simeq& - \frac{i}{2 M} \left ( i \partial \hspace{-6pt}/_x  - g {A \hspace{-6pt}/} + M 
\right ) \exp \left ( \gamma \cdot \frac{\partial}{\partial \Gamma} 
\right ) \> \sum_{s = \pm 1} \int_0^{\infty} dT \>
\exp \left ( -  \frac{i}{2} M T  \right ) \nonumber \\
&& \hspace{1.5cm} \cdot  \, N_4^{\, \rm spin} \, 
\int{\cal D}^4x \> {\cal D}^3\Pi  \> {\cal D}^4\zeta 
\int{\cal D}E \> 
\exp \Bigl \{  \> i S_s[x,{\mbox{\boldmath $\Pi$}},E,\zeta] \> \Bigr \}_{\Gamma = 0} 
\label{nonrel 1}
\end{eqnarray}
with
\begin{eqnarray}
S_s[x,{\mbox{\boldmath $\Pi$}},E,\zeta] &=& - i \zeta(0) \cdot \zeta(T) + \int_0^T dt 
\> \Bigl [ \, \frac{M}{2} - M s \dot x_0 + {\mbox{\boldmath $\Pi$}} \cdot \dot {\bf x} 
- g A_0 \dot x_0 \nonumber \\
&& + g {\bf A} \cdot \dot {\bf x} - \frac{{\mbox{\boldmath $\Pi$}}^2}{2 M} + 
i \zeta \cdot \dot \zeta - \frac{i g}{M} F_{\mu \nu} \zeta^{\mu} 
\zeta^{\nu} + s E  - E \dot x_0 + \frac{E^2}{2 M} \, \Bigr ] \> .
\end{eqnarray}
According to our assumption the last term in the square bracket is 
$ {\cal O} (1/M^3) $ which we neglect. The path integral over $E$ then 
gives a functional $\delta$-function
\begin{equation}
\delta \, \left [ \dot x_0 - s \right ]
    \> = \>\lim_{N \to \infty} \prod_{k=1}^N \> \delta \left( 
\frac{x_{0,k}-x_{0,k-1}}{\Delta t} - s \right) \> , 
\hspace*{1cm} \Delta t=\frac{T}{N}
\end{equation}
The functional integration over $x_0$ can now be performed trivially, 
with the result that the time co-ordinate has the proper time dependence
\begin{equation}
x_0(t) \> = \>y_0 + s \, t  \> .
\label{time vs. prop time}
\end{equation}
However, one $\delta$-function remains because there are only 
$(N-1)$ integrations in the discretized path integral for the 
co-ordinates:
\begin{eqnarray}
G(x,y) &\simeq& - \frac{i}{2 M} \left ( i \partial \hspace{-6pt}/_x  - g {A \hspace{-6pt}/} + M 
\right ) \exp \left ( \gamma \cdot \frac{\partial}{\partial \Gamma} 
\right ) \> \sum_{s = \pm 1} \int_0^{\infty} dT \> \delta \left ( x_0 - 
y_0 - s T \right ) \exp \left ( - i M T  \right ) \nonumber \\
&& \hspace{0.5cm} \cdot  \, N_4^{\, \rm spin} \, 
\int{\cal D}^3x \> {\cal D}^3\Pi  \> {\cal D}^4\zeta \>  
\exp \Biggl \{  \> \zeta(0) \cdot \zeta(T) + i \int_0^T dt \> \Bigl [ 
\, i \zeta \cdot \dot \zeta + {\mbox{\boldmath $\Pi$}} \cdot \dot {\bf x} \nonumber \\
&& \hspace{5cm} - \frac{{\mbox{\boldmath $\Pi$}}^2}{2 M} 
- g s A_0 + g {\bf A} \cdot \dot {\bf x}
- \frac{i g}{M} F_{\mu \nu} \zeta^{\mu} \zeta^{\nu} \, \Bigr ]
\Biggr \}_{\Gamma = 0}  \> .
\label{nonrel 2}
\end{eqnarray}
The remaining $\delta$-function
enforces the boundary condition $x_0(T) = x_0$ and can be used to 
perform the integration over the proper time $T$ yielding
\begin{equation}
T \> = \>s \left ( x_0 - y_0 \right ) \> .
\end{equation}
In other words, in the non-relativistic limit the proper time becomes 
the ordinary time (difference), as expected.
Since the proper time is positive, the $(s = +1)$-term describes
forward propagation of the particle whereas  the $(s = -1)$-term 
describes backward propagation of
the antiparticle, which is also contained in the
Feynman propagator but decouples in the heavy mass limit.
Furthermore, the global projection operator
in front of the propagator (\ref{nonrel 2}) can be replaced by
\begin{equation}
- \frac{i}{2M} \left ( i \partial \hspace{-6pt}/_x  - g {A \hspace{-6pt}/} + M \right ) 
\> \longrightarrow \> - i \, \frac{1}{2} \left ( 1 + s \gamma_0 \right )
+ {\cal O} \left ( \frac{1}{M} \right )
\label{projection}
\end{equation}
as the $x_0$-derivative acting on the phase factor 
$ \exp \left ( - i M s (x_0 - y_0) \right ) $ gives the leading 
contribution. Since
\begin{equation}
\gamma_0 \> = \>
 \left ( \begin{array} {cc}
                                  1        &  0\\
                                  0        &   -1
                     \end{array} \right )
\end{equation}
the (anti-)particle propagator acts only on the 
(lower) upper components of Dirac spinors if the remaining path integral 
is diagonal in $ 2 \times 2 $ Dirac space (which will turn out to be the
case). Shifting back to integration over ${\bf p}$ we therefore obtain
\begin{eqnarray}
G(x,y) &\simeq& - i \sum_{s = \pm 1} \> \Theta \left ( s (x_0 - y_0) 
\right ) \, \frac{1}{2} \left ( 1 + s \gamma_0 \right )
e^{ - i M s ( x_0 - y_0 ) }
\exp \left ( \gamma \cdot \frac{\partial}{\partial \Gamma} \right ) \>
\> N_4^{\, \rm spin} \> \nonumber \\
&& \hspace{1cm} \cdot \int{\cal D}^3x \> {\cal D}^3 p  \> {\cal D}^4\zeta 
\> \exp \Biggl \{  \> \zeta(0) \cdot \zeta(T) + i 
\int\limits_0^{T=s(x_0-y_0)} dt \> \Bigl [
\, i \zeta \cdot \dot \zeta + {\bf p} \cdot \dot {\bf x} \nonumber \\
&& \hspace{4cm} - \frac{1}{2 M} \left ( {\bf p} - g {\bf A} \right )^2 
- g s A_0 - \frac{i g}{M} F_{\mu \nu} \zeta^{\mu} \zeta^{\nu} \, \Bigr ]
\Biggr \}_{\Gamma = 0} \> .
\label{nonrel 3}
\end{eqnarray}
The time-dependence of the electromagnetic potentials and fields is fixed 
by Eq.~(\ref{time vs. prop time}). Substituting
\begin{eqnarray}
t' &=& y_0 + s t \> , \hspace{0.5cm} t' \in [y_0,x_0] \nonumber \\ 
{\bf x}(t) &=& {\bf x}'(t') \> ,
\hspace{0.5cm} {\bf p}(t) \> = \>{\bf p}'(t') \> ,\hspace{0.5cm} \zeta(t) 
\> = \> 
\zeta'(t')
\end{eqnarray}
the boundary conditions for the co-ordinate space path integral
become the usual ones for a nonrelativistic path integral \cite{Schul}
\begin{equation}
{\bf x}'(y_0) \> = \>{\bf y} \> , \hspace{0.3cm} {\bf x}'(x_0) \> = \>{\bf x} \> .
\end{equation}
Omitting the primes, the action in the phase space path integral now 
reads
\begin{eqnarray}
S_s [{\bf x},{\bf p},\zeta] &=& - i \zeta_{\mu}(y_0) \, \zeta^{\mu}(x_0) 
+ \int_{y_0}^{x_0} dt \> \left [ \, i 
\zeta^{\mu} \dot \zeta_{\mu} + {\bf p} \cdot \dot {\bf x} - 
{\cal H}_s ({\bf x},{\bf p},2\zeta) \, \right ] \\
{\cal H}_s ({\bf x},{\bf p},2\zeta) &=& s \left [ M + 
\frac{ ({\bf p} - g {\bf A})^2}{2 M} \right ] + g A_0 + \frac{i g s }{M}
F_{\mu \nu} \zeta^{\mu} \zeta^{\nu} \> .
\end{eqnarray}
Here we have absorbed the phase factor $ \exp (- i M s (x_0 - y_0) ) $ 
into the Hamiltonian ${\cal H}_s$.

Finally we simplify the spin degrees of freedom by using
\begin{equation}
F_{\mu \nu} \zeta^{\mu} \zeta^{\nu} \> = \>2 \zeta_0 \, {\bf E} \cdot
{\mbox{\boldmath $\zeta$}}
 - {\bf B} \cdot \left ( {\mbox{\boldmath $\zeta$}} \times 
{\mbox{\boldmath $\zeta$}} \right )
\label{spin interaction}
\end{equation}
and observing that the first term in Eq.~(\ref{spin interaction})
is linear in $\zeta_0$. After shifting $\zeta_0 = \Gamma_0/2
+ \xi_0 $, the $\xi_0$-integration can be performed and leads to a term
in the remaining action which
is of ${\cal O}(1/M^2)$ : the Fourier transform of a Gaussian is again
a Gaussian. In that process part of the spin normalization factor
is cancelled. The same argument can be applied with respect to the
${\mbox{\boldmath $\zeta$}}$ - integration so that in leading order only
$ \> i g s \Gamma_0 {\bf E } \cdot {\mbox{\boldmath $\Gamma$}}/M  \> $  
survives
from the first term  in Eq.~(\ref{spin interaction}).
Performing the required differentiations
with respect to $ \Gamma_0 $ we see that we obtain a contribution to
Eq.~(\ref{projection}) of the same order which was already neglected.
Notice that $ \> 2 \zeta_0 \, {\bf E} \cdot {\mbox{\boldmath $\zeta$}} \> $
is ``odd'' in the sense of connecting large and small components in the
Dirac equation; from the standard Foldy-Wouthuysen transformation
of the Dirac Hamiltonian it also follows that the odd parts are
suppressed by a factor $1/M$ compared to the ``even'' ones.
In addition, since
\begin{equation}
\gamma_i \gamma_j
\> = \>\left ( \begin{array} {cc}
                                  0        & \sigma_i \\
                            - \sigma_i     &   0
                     \end{array} \right )
\> \left ( \begin{array} {cc}
                                  0        & \sigma_j \\
                            - \sigma_j     &   0
                     \end{array} \right )
\> = \>- \left ( \begin{array} {cc}
                         \sigma_i  & 0 \\
                              0    & \sigma_i
                     \end{array} \right )
\> \left ( \begin{array} {cc}
                         \sigma_j  & 0 \\
                             0     & \sigma_j
                     \end{array} \right )
\end{equation}
one can set 
\begin{equation}
\exp \left ( {\mbox{\boldmath $\gamma$}} \cdot 
{\mbox{\boldmath $\nabla$}}_{\Gamma} \right ) \>
\longrightarrow \> \exp \left ( i {\mbox{\boldmath $\sigma$}} \cdot
{\mbox{\boldmath $\nabla$}}_{\Gamma} \right ) 
\end{equation}
for the remaining even part of the action.  After changing the signs of
 the spin terms by the substitution
$ {\mbox{\boldmath $\zeta$}} \to i {\mbox{\boldmath $\zeta$}} $ the
nonrelativistic limit for the Dirac propagator finally becomes
\begin{eqnarray}
G \left ( {\bf x},x_0; {\bf y},y_0 \right )
&\simeq& - i \sum_{s = \pm 1} \> \Theta \left ( s (x_0 - y_0) \right )
\, \frac{1}{2} \left ( 1 + s \gamma_0 \right ) \> 
e^{{\mbox{\boldmath $\sigma$}} \cdot {\mbox{\boldmath $\nabla$}}_{\Gamma}} \>
\> \bar N_3^{\, \rm spin} \> \nonumber \\
&& \cdot \int\limits_{{\bf x}(y_0)={\bf y}}^{{\bf x}(x_0)={\bf x}}
{\cal D}^3x \> {\cal D}^3 p  \int\limits_{{\mbox{\boldmath $\zeta$}}(y_0)+
{\mbox{\boldmath $\zeta$}}(x_0)={\mbox{\boldmath $\Gamma$}}}
{\cal D}^3 \zeta \> \exp \Biggl \{  \> {\mbox{\boldmath $\zeta$}}(y_0) 
\cdot {\mbox{\boldmath $\zeta$}}(x_0) \nonumber \\
&& \hspace{1.5cm}+ i \int_{y_0}^{x_0} dt \> \Bigl [
\,  i {\mbox{\boldmath $\zeta$}} \cdot \dot {\mbox{\boldmath $\zeta$}} + {\bf p} \cdot \dot {\bf x}
- {\cal H}_s ({\bf x},{\bf p},2 {\mbox{\boldmath $\zeta$}}) \, \Bigr ] \> 
\Biggr \}_{{\mbox{\boldmath $\Gamma$}} = 0} 
\label{nonrel 4}
\end{eqnarray}
with
\begin{equation}
\bar N^{\, \rm spin}_3 \> = \>\left \{ \> \int\limits_{{\mbox{\boldmath $\zeta$}}(y_0)+
{\mbox{\boldmath $\zeta$}}(x_0)={\mbox{\boldmath $\Gamma$}}} {\cal D}^3 \zeta \>
\exp \left [ \, {\mbox{\boldmath $\zeta$}}(y_0) \cdot {\mbox{\boldmath $\zeta$}}(x_0) -
\int_{y_0}^{x_0} dt \> {\mbox{\boldmath $\zeta$}} \cdot \dot {\mbox{\boldmath $\zeta$}} \,
\right ] \>
\right \}^{-1} \> .
\label{N nonrel spin}
\end{equation}
The Hamiltonian
\begin{equation}
{\cal H}_s ({\bf x},{\bf p},2 {\mbox{\boldmath $\zeta$}}) \> = \> s \left [ M +
\frac{ ({\bf p} - g {\bf A})^2}{2 M} \right ] + g A_0 + \frac{i g s }{M}
{\bf B} \cdot \left ( {\mbox{\boldmath $\zeta$}} \times {\mbox{\boldmath $\zeta$}} \right ) 
\end{equation}
coincides exactly with the standard Foldy-Wouthuysen Hamiltonian
for particles and antiparticles
(see, e.g. Ref.~\cite{BjDr}, Eq.~(4.5) ) 
\begin{equation}
H_{\rm FW} ({\bf x},{\bf p},{\mbox{\boldmath $\sigma$}})
 \> = \>\gamma_0 \, \left [ M +
\frac{ ({\bf p} - g {\bf A})^2}{2 M} \right ] + g A_0 - \frac{g}{2 M}
\gamma_0 \, {\mbox{\boldmath $\sigma$}} \cdot {\bf B} + 
{\cal O} \left ( \frac{1}{M^2} \right )
\label{FW Hamiltonian}
\end{equation}
if the last term (the so-called Pauli term) is rewritten using 
\begin{equation}
{\mbox{\boldmath $\sigma$}} \times {\mbox{\boldmath $\sigma$}}\> = \>2 i \, {\mbox{\boldmath $\sigma$}} \> .
\label{sigma ident}
\end{equation}
This ensures that the time evolution is governed by a Grassmann-even
Hamiltonian. 
It is, of course, straightforward to
start from the nonrelativistic spin-dependent Hamiltonian
(\ref{FW Hamiltonian}) and by using the identity (\ref{sigma ident})
to derive the 3-dimensional path integral
(\ref{nonrel 4}) for the propagator as has been done in Ref.~\cite{AFP},
Section~5. Here we proceeded in the reverse order showing 
how the nonrelativistic limit can be taken within the path integral 
representation of the Dirac propagator.
It should also be possible to evaluate
higher-order terms in the nonrelativistic reduction in this way or to
obtain the semiclassical limit of the propagator \cite{BoKe}.

\section{Dirac Propagator in an External Scalar Potential}

For some applications one needs the propagator of a fermion which moves 
in an external scalar field $S(x)$ as well. For example, in the 
Walecka model \cite{Wal}
the exchange of a scalar meson generates attraction between 
nucleons whereas massive vector mesons are responsible for 
repulsion at shorter distances. In such cases we need to evaluate the 
following Green function
\begin{equation}
G(x,y) \> = \>\left < x \, \left | \, \frac{1}{{\Pi \hspace{-8pt}/} - M^{\star}(x) + i 0 } 
\, \right | \, y \right > 
\label{G scalar}
\end{equation}
where 
\begin{equation}
M^{\star}(x) \> = \>M + S(x)
\end{equation}
is the effective, position-dependent mass of the fermion.

The previous method of multiplying numerator and denominator in 
Eq.~(\ref{G scalar}) by $ {\Pi \hspace{-8pt}/} + M^{\star}$ obviously 
does not work anymore since
\begin{equation}
\left ( {\Pi \hspace{-8pt}/} - M^{\star} \right ) \left ( {\Pi \hspace{-8pt}/} + M^{\star}
\right ) \> = \>{\Pi \hspace{-8pt}/}^2 - M^{\star \> 2} + \left [ {\Pi \hspace{-8pt}/}, M^{\star}
\right ]
\end{equation}
is not Grassmann even. Consequently there are statements in the 
literature \cite{RSchSch}
that in this case a five-dimensional formalism is the only possible 
approach. However, this is {\it not} the case: the problem to rationalize 
the denominator of the Green function is analogous to the problem of
inverting complex matrices by using only real arithmetic. This is easily
achieved by writing
\begin{equation}
\frac{1}{ A + i B} \> = \>\left ( \> 1 - i A^{-1} B \> \right ) \>  
\frac{1}{ A + B A^{-1} B } \>.
\end{equation}
Therefore we have
\begin{equation}
\frac{1}{ {\Pi \hspace{-8pt}/} - M^{\star} } \> = \>\left ( M + \frac{M}{M^{\star}}
{\Pi \hspace{-8pt}/} \right ) \> \frac{1}{{\Pi \hspace{-8pt}/} \frac{M}{M^{\star}}{\Pi \hspace{-8pt}/} - M 
M^{\star} + i 0 }
\label{Ravndal scalar}
\end{equation}
as the Di Vecchia-Ravndal representation for the present case. Again we 
have the choice to project on $ {\Pi \hspace{-8pt}/} = M^{\star}$ either during the 
evolution or at the end. If we adopt the latter approach the quantum 
mechanical Hamiltonian which governs the proper time evolution is now
\begin{equation}
\hat {\cal H} \> = \>- \frac{1}{2 \kappa_0} \hat {\Pi \hspace{-8pt}/} 
\frac{M}{M^{\star}(\hat x) } \hat {\Pi \hspace{-8pt}/} \> .
\label{H with scalar pot}
\end{equation}
The phase-space path integral representation of the propagator is now 
determined by evaluating the Wigner transform of 
Eq.~(\ref{H with scalar pot}) (see Appendix A). With the abbreviation
$ \> U(x) = M/M^{\star}(x) \> $ one obtains
\begin{eqnarray}
{\cal H}(p,x,\gamma) &=& - \frac{1}{2 \kappa_0} \Pi^2 \, U(x)
+ \frac{i g}{4 \kappa_0} \gamma_{\mu} \gamma_{\nu}
F^{\mu \nu}(x) \, U(x) \nonumber \\
&& - \frac{i g}{4 \kappa_0} \gamma_{\mu} \gamma_{\nu} \left [ \, 
\partial^{\mu} U(x) \, \Pi^{\nu} - \Pi^{\mu} \partial^{\nu} U(x) \, 
\right ] -  \frac{1}{8 \kappa_0} \partial^2 U(x)
\> .
\end{eqnarray}
Since the Hamiltonian is quadratic in $ \Pi = p - g A$ the momentum 
path integral can  still be performed so that the Lagrangian path 
integral representation of the propagator reads
\begin{eqnarray}
G(x,y) \! \! &=& \! \! - \frac{i}{2 \kappa_0} \,
\Bigl [  \, U(x) \Bigl ( i \partial \hspace{-6pt}/ - g {A \hspace{-6pt}/}(x) \Bigr ) + M  \, \Bigr ]
\int_0^{\infty} dT  \, N(T) \, \exp \left (  \, - 
\frac{i M M^{\star}(x) T}{2 \kappa_0} 
\, \right ) \, \exp \left ( \gamma \cdot \frac{\partial}{\partial \Gamma}
\right ) \nonumber \\
&& \hspace{1cm} \cdot \int_{x(0)=y}^{x(T)=x} {\cal D}x 
\int\limits_{\zeta(0)+\zeta(T)=\Gamma}
{\cal D}\zeta \> \exp \left \{ \, \zeta(0) \cdot \zeta(T) + 
i \int_0^T dt \> L(x,\dot x,\zeta,\dot \zeta) \, \right \}_{\Gamma=0}
\end{eqnarray}
with
\begin{eqnarray}
L(x,\dot x,\zeta,\dot \zeta) &=& i \zeta \cdot \dot \zeta - 
\frac{\kappa_0}{2 U(x)} \dot x^2 - g A(x) \cdot \dot x - 
\frac{ig}{\kappa_0} U(x) F^{\mu \nu}(x) \zeta_{\mu} \zeta_{\nu} 
\nonumber \\
&& - 2i \dot x \cdot \zeta \frac{1}{U(x)} \zeta \cdot \partial U(x) + 
\frac{1}{8 \kappa_0} \partial^2 U(x) + 2 i \delta(0) \log U(x) \> .
\label{Lagr S+V}
\end{eqnarray}
The last term arises from the quadratic fluctuations 
\begin{equation}
\prod_k \frac{1}{U^2(x_k)} \> = \>\exp \left ( - 2 \sum_k \log U(x_k) 
\right ) \> = \>\exp \left ( i  \, 2 i \, \frac{1}{\Delta t} \, \Delta t 
\sum_k \log U(x_k) \right )
\end{equation}
in the discretized momentum path 
integral which are now position-dependent due to the effective mass 
$M^{\star}(x)$. The awkward $\delta(0)$ appears as the formal limit of 
$ 1/\Delta t$ when the time-slicing $ \Delta t $ is made infinitesimal
(see Ref.~\cite{Lee}, Chapter 19) and cancels consistently against other
divergencies \cite{delta(0)}.

\section{Effective Action for Quenched QED}

In order to reduce the number of degrees of freedom it is
advantageous for some applications to integrate out the bosons which 
mediate the interactions. The price to be paid is, of course, a more 
complicated two-time effective interaction. We will outline this 
procedure by 
considering Quantum Electrodynamics (QED) (or the Walecka model without 
scalar mesons~\footnote{If one also wants to integrate out the
scalar mesons, the 5-dimensional Berezin-Marinov description has
to be used because only then a Gaussian path integral for the scalar mesons
is obtained; our 4-dimensional form (\ref{Lagr S+V}) is highly nonlinear in
$ S(x) $ .})
\begin{equation}
{\cal L} = {\cal L}_0 (A) + \bar{\psi} \left ( i\partial \hspace{-6pt}/-g{A \hspace{-6pt}/}-M_0 
\right ) \psi  
\end{equation}
where 
\begin{equation}
{\cal L}_0 (A)  \> = \>-\frac{1}{4} F_{\mu\nu} F^{\mu\nu} + \frac{1}{2} m^2 
A^2 -\frac{1}{2} \lambda (\partial \cdot A)^2
\end{equation}
is the St\"uckelberg Lagrangian with a gauge parameter $\lambda$.
We have given the photons a mass $m$ in order to regularize infrared 
divergencies.

The generating functional for the 2-point function with an 
arbitrary number of photons  is  
\begin{equation}
Z'\> [j,x] \> = \>\int {\cal D}A \>  \left < x\, \left | \, 
\frac{1}{i\partial \hspace{-6pt}/-g{A \hspace{-6pt}/}-M_0} \, \right |\, 0 \right > 
\> \exp \Bigl \{ i {\cal A}_0 [ A ] + (j,A) \> \Bigr \} \> .
\label{gen' functional}
\end{equation}
Here the free vector meson action is denoted by
$ {\cal A}_0 [ A ] = \int d^4 x \> {\cal L}_0 (A) $ and we have 
neglected closed fermion loops (quenched approximation) \cite{WC1}
in order to have a single world-line for the fermion.
For integrating out the vector field $A_{\mu}$ we
use the path integral representation of the Dirac propagator in an 
external vector field in its supersymmetric form and 
the gaussian integration formula
\begin{equation}
\int {\cal D}A_{\mu} \> \exp \left [ \frac{i}{2} \left ( A_{\mu}, 
(G^{-1})^{\mu\nu}
A_{\nu} \right ) + \left ( A_{\mu},h^{\mu} \right ) \> \right ] \>
\propto \> \exp \left [ \frac{i}{2} \left ( h_{\mu}, G^{\mu\nu}
h_{\nu} \right ) \> \right ] \> .
\end{equation}
Here
\begin{equation}
G^{\mu\nu}(k) \> = \> 
-\left[\frac{g^{\mu\nu}- k^{\mu} 
k^{\nu}/m^2}{k^2-m^2} + \frac{k^{\mu} k^{\nu}/m^2}{k^2-m^2/\lambda}
\right]
\label{photon propagator}
\end{equation}
is the standard propagator for massive vector particles 
(Ref.~\cite{ItZu}). From the linear terms in $A^{\mu}$ we read off
\begin{equation}
h_{\mu}(y) \> = \>j_{\mu}(y) \> + \> i g \int_0^T dt \int d\theta
\> D X_{\mu}(t,\theta) \> \delta^4 ( \> y - X(t,\theta) \> ) \> .
\label{def h}
\end{equation}

For the present purposes the local projection method is preferable 
because the whole dependence
on the photon field resides in the free photon action and the
electron-photon interaction.
After integration over the photon field we then obtain
for the generating functional (\ref{gen' functional})
\begin{eqnarray}
Z'\> [j,x] &=& {\rm const.} \> 
\exp \left ( \gamma \cdot \frac{\partial}{\partial \Gamma} \right ) 
\int_0^{\infty} dT \> N(T) \exp \left ( - \frac{i}{2 \kappa_0}  M^2 T
 \right ) \nonumber \\
&& \hspace{2cm} \cdot \int d\chi \>
\exp \left ( - \frac{i}{2 \kappa_0}  M \chi  \right ) 
{\cal D}x {\cal D}\zeta  \> e^{ \> i S_{\rm eff} [X,j] } \> \>
\Biggr|_{\Gamma=0} 
\label{eq: genfunc}
\end{eqnarray}
where the effective action is given by
\begin{equation}
S_{\rm eff} [X,j] \> = \>S_0 \, '[X]
\> + \> \frac{1}{2} \left ( h_{\mu}, G^{\mu\nu} h_{\nu} \right ) \> .
\end{equation}
As in Ref.~\cite{WC1}, it is advantageous to split it up into
terms involving zero, one or two external sources $j(y)$. The latter one
leads to disconnected diagrams and can be discarded. We then have
\begin{equation}
S_{\rm eff} [X,j] \> = \>S_0 \> '[X] \> \> + \> S_1[X] \> + \> S_2[X,j] 
\label{eq: seff}
\end{equation}
where the free action is given in Eq.~(\ref{susy S0'}) and the 
interaction part by
\begin{eqnarray}
S_1[X] &=&  \frac{g^2}{2} \int_0^T dt_1 \int d\theta_1 \> \int_0^T dt_2
\int d\theta_2 \> \int \frac{d^4 k}{(2 \pi)^4}
 \> G^{\mu\nu}(k)  \> D X_{\mu} (t_1,\theta_1) \>
D X_{\nu} (t_2,\theta_2) \nonumber \\
&& \hspace{5.5cm} \cdot \> \exp \left \{ \> - i k \cdot \left [ \, 
X(t_1,\theta_1) - X(t_2,\theta_2) \, \right ] \> \right \} \> .
\label{susy S1} 
\end{eqnarray}
Note that the ``current''
\begin{equation}
J_{\mu} (X) \> \equiv \> D X_{\mu}(t,\theta) \> = \>- \theta 
\dot x_{\mu}(t) + a \> \zeta_{\mu}(t)
\end{equation}
looks like in scalar QED but is Grassmann odd and does not depend on the
integration variable $k$. Therefore the $k$-integration can be performed
easily giving the photon propagator in configuration space with argument
$X(t_1,\theta_1) - X(t_2,\theta_2)$~\cite{SchSch}.
Written in components the interaction term
\begin{eqnarray}
S_1[x,\zeta] &=& - \frac{g^2}{2} \int_0^T dt_1 dt_2 \int
\frac{d^4 k}{(2 \pi)^4}
 \> G^{\mu\nu}(k)  \>
\left [ \, \dot x_{\mu}(t_1) \, + \,
\frac{2}{\kappa_0} \zeta_{\mu}(t_1) \, k \cdot \zeta(t_1) \, \right ]
\nonumber \\
&& \hspace{3cm} \cdot \left [ \, \dot x_{\nu}(t_2) \, - \,
\frac{2}{\kappa_0} \zeta_{\nu}(t_2) \> k \cdot \zeta(t_2) \, \right ]
\> e^{- i k \cdot \left [ \, x(t_1) - x(t_2) \, \right ] }
\end{eqnarray}
is seen to contain up to {\it quartic} terms in the spin variable $\zeta$.
This means that, unlike the case of external fields, the Grassmann variables
cannot be integrated out anymore to give a ``spin factor''. Vice versa, 
it is impossible to eliminate the photon field starting from the spin 
factor formulation for the propagator.

\noindent
The source term becomes
\begin{equation}
S_2[X,j] \> = \>i g \int d^4 y \> j_{\mu}(y) \int_0^T dt \, \int d\theta 
\> \int \frac{d^4 k}{(2 \pi)^4} \> G^{\mu\nu}(k) \>
\> D X_{\nu} (t,\theta) \> \exp \left \{ \,  - i k \cdot \left [ 
X(t,\theta) - y \right ] \, \right \} .
\label{susy S2}
\end{equation}
It is also possible to use the global projection method which does not 
have a spin-orbit coupling. However, there is an additional dependence on 
the photon field in the covariant derivative acting on the path integral 
in Eq.~(\ref{G propagator}) which makes it less suitable for deriving an 
effective action.

To conclude this Section, we note that the effective action in
Eq.~(\ref{eq: seff}) allows a particularly concise derivation of the
transformation properties of Green functions 
\footnote{These transformations were first derived
for the electron propagator and the electron-photon vertex
by Landau and Khalatnikov~\cite{LaKh} and extended to general Green
functions by Fradkin and Zumino~\cite{FradZum}. }
under a change of the gauge parameter $\lambda \> $ :
We see from the photon propagator $G_{\mu \nu}(k)$ in 
Eq.~(\ref{photon propagator}) that a change in $\lambda$ only effects 
the term proportional to $k_{\mu} k_{\nu}$.  For this term the integrals 
over the proper times $t_i$ and $\theta_i$ occurring in the effective 
action may be performed exactly as the integrand is a total derivative, 
i.e.
\begin{equation}
\int_0^T   dt \> \int d\theta \> k\cdot DX(t,\theta) \;
e^{-i k \cdot X(t,\theta)} \> = \> 
i \> \int_0^T dt \> \int d\theta \> D \, e^{-i k \cdot X(t,\theta)}
\> = \>  i \left ( 1 - e^{-i k\cdot x}  \right )\;\;\;.
\end{equation}
The change in $S_1$ (Eq.~(\ref{susy S1})) and $S_2$ (Eq.~(\ref{susy S2})) 
induced by a change in $\lambda$ from $\lambda_1$ to $\lambda_2$, say, 
is therefore only dependent on the endpoints of the path $x(t)$ and 
not on the path itself. If we define $\Delta (x^2)$ to be the Fourier 
transform of the change of the coefficient  ($\equiv \tilde \Delta (k^2)$) 
of $k_{\mu} k_{\nu}$ in the photon propagator, 
i.e.
\begin{equation}
\Delta (x^2) \> = \>-{1 \over m^2} \>
  \int \frac{d^4 k}{(2 \pi)^4} \> 
\left (\frac{1}{k^2-m^2/\lambda_2} \> - \>\frac{1}{k^2-m^2/\lambda_1} 
\right ) \> e^{-i k \cdot x} \;\;\;,
\end{equation}
then the corresponding change in $S_1$ is given by 
\begin{eqnarray}
\delta S_1 &=& S_1^{\lambda=\lambda_2}[X] \> -\> 
S_1^{\lambda=\lambda_1}[X] \> = \> {g^2 \over 2} \> \int 
\frac{d^4 k}{(2 \pi)^4} \> \tilde \Delta (k^2) \>i^2 \>
\left ( 1 - e^{-i k\cdot x}  \right ) \left ( 1 - e^{i k\cdot x}  
\right )\nonumber \\
&=& g^2 \left [\Delta(x^2) \> - \> \Delta(0) \right ]
\end{eqnarray}
while the change in $S_2$ is
\begin{equation}
\delta S_2 \> = \> -i g \> \int d^4 y \> j(y) \cdot \partial_y
\left[ \, \Delta \left ([y-x]^2 \right ) \> - \> \Delta(y^2) \, 
\right ]\;\;\;.
\end{equation}
Note that not only is the change in $S_{1,2}$ independent of the path, so
that it may be pulled out of the path integral in Eq.~(\ref{eq: genfunc}), 
it also does not involve the Grassmann valued $\Gamma$ nor is it dependent 
on the proper time $T$. 
Hence the generating function for the Green functions with gauge parameter 
$\lambda_2$ is related to that with gauge parameter 
$\lambda_1$ in a very simple way, namely
\begin{eqnarray}
Z'_{\lambda_2} [j,x] &=& 
e^{i \left (\delta S_1 + \delta S_2 \right ) } \> Z'_{\lambda_1} [j,x] \\
&=& \exp \left \{\> i g^2 \left [\, \Delta(x^2) - \Delta(0) \, \right ] +   
g \int d^4 y \> j(y) \cdot \partial_y
\left[\, \Delta \left ([y-x]^2 \right ) - \Delta(y^2) \, \right ] \> 
\right \} \> Z'_{\lambda_1} [j,x] .\nonumber
\end{eqnarray}
As special cases we can derive the transformation laws for the
propagator and the electron-photon vertex from this expression 
\footnote{See Ref.~\cite{LaKh}.  Note that in that paper the photon 
propagator is defined with a minus sign with respect to ours.  Hence our
function $\Delta(y^2)$ is $-\Delta_F(y)$ of Ref.~\cite{LaKh} and our
untruncated vertex function is the negative of the function $B_\mu$
defined by Landau {\it et al.}}.  Setting $j=0$ we obtain
\begin{equation}
G^{\lambda_2}(x,0) \> = \> e^{i g^2 \left [\Delta(x^2) \> - \> \Delta(0) 
\right ] } \> G^{\lambda_1}(x,0) \;\;\;,
\end{equation}
while by differentiating once with respect to the current and then setting 
$j=0$ we obtain the (untruncated) vertex function
\begin{equation}
G_{2,1}^{\lambda_2 \> \mu}(y;x,0) \> =\> g \left \{ \partial_y^\mu
\left[\Delta([y-x]^2) \> - \> \Delta(y^2) \right ] \right \} 
G^{\lambda_2}(x,0) 
\> + \> e^{i g^2 \left [\Delta(x^2) \> - \> \Delta(0) \right ] }
\>G_{2,1}^{\lambda_1 \> \mu}(y;x,0) \;\;\;.
\end{equation}
It should be noted that these relations are valid even if the 
photon mass (which violates gauge invariance) is kept non-zero
in the photon propagator.

\section{Summary and Conclusions}

The main purpose of this work is to explore a four-dimensional path
integral representation for the Dirac propagator 
 general enough to describe the particle's motion
in  both vector and scalar fields.    
Although the four-dimensional approach, for an external vector field,
was proposed by  Di Vecchia and Ravndal almost twenty years ago, it 
had received limited attention up to now.  Instead  it is standard
to use the Berezin - Marinov approach where one introduces a fifth 
component to eliminate  the  extra spin degree of freedom. However, 
the fifth component has no clear physical meaning and the necessity of 
introducing a $\gamma_5$ for the evaluation of the propagator seems 
rather unnatural. The  four-dimensional representation avoids these 
difficulties; in addition the supersymmetry transformations become 
easier and more natural to  generate.
 
Working within this four-dimensional formalism we have presented two
alternative methods to project out the unwanted spin degree of freedom.
The first method  projects onto the final state after the time evolution
and is hence termed global, whereas in the second method the projection
is done at each step in the time 
evolution and it is therefore referred to as local. Extending previous work
by Reuter, Schmidt and Schubert we have shown
that due to the supersymmetry the two methods are completely equivalent
and may be used according to convenience. The main difference between the
two approaches is that the path integral representation using the local
projection has an explicit spin-orbit coupling term. It was therefore
crucial for the proof of equivalence to generalize the results of Ravndal
and Di Vecchia regarding the supersymmetry transformations to apply also 
in the case where the spin-orbit term appears. In Refs.~\cite{DiVR,Rav} 
it was pointed out that, in the case where no spin-orbit term was
present, a simple way of generating both bosonic and fermionic 
transformations is to shift the times $t$ and $\theta$. 
We show in this paper that in the presence of a spin-orbit
term in addition to a shift an appropriate scaling of the times $t$ 
and $\theta$
is needed in order to generate the correct supersymmetry transformations.
This scaling is such that the parameter $\kappa_0$ and the supertime 
$\chi$ remain unchanged.    

For the case of a Dirac particle in an external scalar
potential it was generally believed that a five-dimensional
approach was unavoidable. We have here shown that this
is {\it not} the case and we used the four-dimensional 
description to obtain the Dirac propagator in an external scalar field~.  

Despite the attention given to  spin in the path integrals, 
a nonrelativistic reduction starting directly from the Dirac propagator 
was still missing. By expanding the relativistic expression 
in powers of $1/M$ we were able to reduce the 
path integrals to three-dimensional form and to obtain 
the leading nonrelativistic 
result described by the Foldy-Wouthuysen Hamiltonian.  

Finally we applied the four-dimensional approach to quenched QED in
order to obtain a supersymmetric formulation for the generating
functional of Green functions with one electron line and an arbitrary
number of external photon lines.  It was possible to do this as in the
quenched approximation the photons can be integrated out, yielding a
path integral only in the electron degrees of freedom, albeit with a
complicated nonlocal interaction. In this form one can apply methods
along the same lines as those used in the study of the polaron problem
as described in Ref.~\cite{ARS}.  Furthermore, we showed that it is a 
rather simple matter to derive the Landau-Khalatnikov transformations
for the propagator, vertex function and indeed any higher-point
function from this formalism.

From a field-theoretic point of view, the worldline technique is
particularly appropriate whenever one deals with a situation where
internal fermion loops may either be neglected or taken into account
perturbatively.  As this situation arises quite naturally in the
non-relativistic regime, the technique would appear to be particularly
appropriate in that setting. We think that the reason it has not
received a great deal of attention by physicists working in that area
is partly due to the fact that the commonly used five-dimensional
representation appears artificial within this context.  In this paper
we have tried to convey the message that for most problems the
five-dimensional formulation is not only unnecessary but in fact
less transparent than the four-dimensional one.  It is our hope,
therefore, that this paper makes worldline techniques more accessible
to a wider audience than they have been up to now.

\noindent  {\it Note added in Proof:}  Recently it was
      pointed out to us that the elimination of the fifth 
      spin variable was also considered by T. Allen
      using Hamiltonian methods (T. Allen, Phys. Lett. B 214, 87 (1988);
      see also T. Allen, Caltech Ph.D. Thesis, (1988) ).  Also,
      J. W. van Holten has advocated the use of a commuting rather
      than anti-commuting fifth spin variable and a different 
      four-dimensional approach (see the first
      reference in [10] as well as a more concise discussion of the
      problem in Nuc. Phys. B (Proc. Suppl.) 49, 319 (1996) ).  
      Finally, another important contribution to the literature on spin
      in path integrals missing from Ref. [6] is the paper by
      M. Halpern, A. Jevicki and P. Senjanovic, (Phys. Rev. D 16, 
      2476 (1977) ).  We are grateful to Profs. T. Allen, M. Halpern 
      and J. W. van Holten for correpondence regarding these
      references.

\begin{acknowledgements}
We would like to thank Michael Marinov for helpful discussions.
One of us (AWS) is supported by the Australian Research 
Council through an Australian Research Fellowship.

\end{acknowledgements}

\appendix

\section*{Spin Path Integral for the Time Evolution Operator}

Here we consider the matrix element of the time evolution operator
\begin{equation}
U(x,y) \> = \>\left < x \, \left | \> \exp \left ( - i \hat {\cal H} \> T 
\right ) \, \right | \, y \right >
\end{equation}
where
\begin{equation}
\hat {\cal H} \> = \>{\cal H} (\hat p, \hat x, \gamma)
\end{equation}
is a Weyl-ordered Hamiltonian. 
Breaking up the evolution operator in $N$ time steps we obtain in the 
usual way
\begin{eqnarray}
 U(x,y) &=&  \lim_{N \to \infty}
\int d^4 x_1 ... d^4 x_{N-1} \> \frac{d^4 p_1}{(2 \pi)^4} ...
\frac{d^4 p_N}{(2 \pi)^4} \> \exp \left [ - i \sum_{i=1}^N p_i \cdot
(x_i - x_{i-1} ) \right ] \nonumber \\
&& \hspace{1cm} \cdot \> \exp
\left [ \> - i  {\cal H}_W(p_N,x_N,\gamma_N) \Delta t \> \right ] \> 
\ldots \> \cdot \> \exp \left [\>  - i  {\cal H}_W(p_1,x_1,\gamma_1) 
\Delta t \> \right ]
\label{time-sliced phase space PI}
\end{eqnarray}
with $ x_0 = y $ and $ x_N = x $. Here 
\begin{equation}
{\cal H}_W(p,x,\gamma) \> = \>\int d^4 y \> \left < x -\frac{y}{2} \> 
\left | \> \hat {\cal H} \> \right | \> 
x + \frac{y}{2} \right > \, e^{-i p \cdot y}
\end{equation}
is the Wigner transform (or Weyl symbol) of the Hamiltonian which is the 
closest classical analogue to the (Weyl-ordered) quantum operator 
\cite{Wig}. We will suppress the subscript ``W'' in the following.

\noindent
There are two essential steps to derive a path integral with spin:
\begin{itemize}
\item[(i)]
Because the Dirac matrices do not commute
the ordering of the factors is essential and the exponentials cannot be
combined with impunity. As is well known this also happens in ordinary
quantum mechanics for time-dependent Hamiltonians.
We therefore have assigned an artificial time-dependence to the Dirac 
matrices and can write now the time evolution operator as a time-ordered 
path integral~\cite{Schul}
\begin{eqnarray}
 U(x,y) &=&  
\int{\cal D}x {\cal D}p \> \> {\cal T}
\exp \left \{ - i \int_0^T dt \left [ \> p \cdot \dot x  +
{\cal H} \left (p(t),x(t),\gamma(t)\right ) \> \right ] \> \right \}
\nonumber \\
&=&  
\int{\cal D}x {\cal D}p
\> \exp \left \{ - i \int_0^T dt \left [ \> p \cdot \dot x  +
{\cal H}\left ( p(t),x(t),\frac{\delta}{\delta \rho(t)} \right ) \>
\right ] \> \right \}
\nonumber \\
&& \hspace{5cm} \cdot \> {\cal T} \exp \left [ \int_0^T dt \> 
\rho^{\mu}(t) \gamma_{\mu}(t) \right ]_{\rho^{\mu}=0} \> .
\label{G hat with diff}
\end{eqnarray}
Here $ \rho^{\mu}(t) $ are Grassmann sources which are assumed to 
anticommute with the Dirac matrices.
The boundary conditions for the $x$-space path integral are
\begin{equation}
x_{\mu}(0) \> = \>y_{\mu} \> ,\hspace{2cm} x_{\mu}(T) \> = \>x_{\mu}  \> .
\label{boundary cond x}
\end{equation}
The time-ordering symbol $ {\cal T} $ would be disastrous for further
manipulation of the path integral. However, in the special case it
can be eliminated by the relation
\begin{eqnarray}
V(T) &\equiv& {\cal T} \exp \left \{ \int_0^T dt \> \rho^{\mu}(t)
\gamma_{\mu}(t) \>
\right \}  \nonumber \\
&=& \exp \left \{-\int_0^T dt_1 \int_0^{t_1} dt_2 \> \rho^{\mu}(t_1)
\rho_{\mu}(t_2)\> \right \} \> \cdot
        \> \exp \left \{\int_0^T dt \> \rho^{\mu}(t)\gamma_{\mu} \> 
\right \}  \> .
\label{time ordering eliminated}
\end{eqnarray}
This can be proved by solving the corresponding evolution equation
\begin{equation}
\frac{\partial  V(T)}{\partial  T} \> = \>\rho^{\mu}(T)\gamma_{\mu} \> 
V(T) \> \> , \hspace{0.5cm} V(0) \> = \>1
\end{equation}
by using the Magnus expansion \cite{Wil}
\begin{equation}
V(T) \> = \>\exp \Biggl \{ \>  \int_0^T dt \> \rho^{\mu}(t) \gamma_{\mu}
\> + \> \frac{1}{2} \int_0^T dt_1 \int_0^{t_1} dt_2 \>
\left [ \rho_{\mu}(t_1) \gamma^{\mu} , \rho_{\nu}(t_2) \gamma^{\nu} 
\right ] \> + \ldots \> \Biggr \} \> .
\end{equation}
The commutator yields $ - 2 \rho_{\mu}(t_1) \rho^{\mu}(t_2) $ which is a 
commuting  c-number so that all higher terms in the expansion which 
involve multiple commutators vanish . On the right-hand-side of
Eq.~(\ref{time ordering eliminated}) we can now drop the artificial
time-dependence of the Dirac matrices.

\item[(ii)]
The differentiations with respect to
$\rho^{\mu}(t)$ which are required in Eq.~(\ref{G hat with diff})
can only be performed easily if they appear linearly in the
exponent. This can be achieved by ``undoing the square'', which is a 
standard procedure \cite{Schul}. However, because $\rho^{\mu}(t)$ is 
anticommuting and one needs an even object in the exponent as evolution 
operator, we have to do it with the help of
a Grassmann path integral. We thus use the identity
\begin{eqnarray}
\exp \left \{- \int_0^T \! dt_1 \int_0^{t_1} \! dt_2 \rho^{\mu}(t_1)
\rho_{\mu}(t_2) \right \} \! &=& \!
\int {\cal D}\xi \> \exp \left \{\int_0^T
dt \left[ - \xi_{\mu}(t)\dot{\xi}^{\mu}(t) +
2 \rho^{\mu}(t)\xi_{\mu}(t)\right] \right \}
\nonumber \\
&&  \cdot \>
\left [ \int {\cal D} \xi \>\> \exp \left ( - \int_0^T
dt \> \xi_{\mu}(t)\dot{\xi}^{\mu}(t)\right ) \right ]^{-1} \> .
\label{Berezin integral}
\end{eqnarray}
and the  antiperiodic boundary condition 
$\> \xi_{\mu}(0) + \xi_{\mu}(T) = 0 \> $ for the Grassmann path integral. 
The standard way of proving this identity in the continuum formulation
is by solving the (differential) equations of motion 
which should give the exact result for quadratic 
actions. However, it is very useful (and reassuring) to have an 
unambiguous formulation with finite time steps $\Delta t $, which 
we will present now: the discretized form of 
\begin{equation}
S \> = \>- \int_0^T
dt \> \left[ - \xi_{\mu}(t)\dot{\xi}^{\mu}(t) +
2 \rho^{\mu}(t)\xi_{\mu}(t)\right]
\end{equation}
may be written as
\begin{equation}
S \> = \>\sum_{i=1}^{N}\left [ 
          \xi_{i,\mu}\left ( \frac{\xi_{i+1}^\mu-\xi_{i-1}^\mu}{2} 
\right)  -{\Delta t \over 2}\>\rho_{i,\mu} \left(\xi_{i+1}^\mu+2\xi_i^\mu
+\xi_{i-1}^\mu \right) \right ] \quad,
\label{S discrete}
\end{equation}
where $ \Delta t = T/N $ and $N$ needs to be even for
the path integral to be an even quantity.
In discretized form the path integral 
over $\xi$ in Eq.~(\ref{Berezin integral}) can now be done by the stationary 
phase method. The (difference) equation of motion 
\begin{equation}
\xi_{k+1}^\mu-\xi_{k-1}^\mu \> = \>- \frac{\Delta t}{2} \, \left( \, 
\rho_{k+1}^{\mu} + 2\rho_k^{\mu} + \rho_{k-1}^{\mu} \, \right)
\label{xi equation of motion}
\end{equation}
can be solved using antiperiodic boundary conditions for $\xi$,
 i.e. $\xi_{N}^\mu$ = $-\xi_{0}^\mu$,
$\xi_{N+1}^\mu$ = $-\xi_{1}^\mu$.  It is convenient (but not necessary)
to impose the equivalent boundary conditions for $\rho$.
Note that the particular discretization of $\rho(t) \cdot \xi(t)$
in Eq.~(\ref{S discrete}) is chosen so that the equations of motion
(\ref{xi equation of motion}) for the odd and even sites are 
coupled.  This avoids  the infamous
``fermion doubling'' problem.  The solution to the equation of motion
is
\begin{equation}
\xi_{{\rm cl}\> j}^\mu \> = \> \frac{\Delta t}{2} \rho_j^\mu 
                - \Delta t \sum_{k=1}^{j} \rho_k^{\mu} + 
            \frac{\Delta t}{2} \sum_{k=1}^{N} \rho_k^{\mu}  \> . 
\label{classical solution} 
\end{equation}
Substituting the solution $\xi_{{\rm cl}}$ 
into Eq.~(\ref{S discrete}) we find
\begin{equation}
S_{\rm cl} =  (\Delta t)^2\sum_{i=1}^N\rho_{\mu,i}\sum_{k=1}^{i-1}
\rho_k^\mu  +\frac{(\Delta t)^2}{8}\sum_{i=1}^N\rho_{\mu,i}
\left (\rho_{i+1}^\mu-\rho_{i-1}^{\mu} \right )\> .
\label{S classic}
\end{equation}
The path integral over $\xi$ can now be performed yielding
\begin{equation}
\frac{\int {\cal D}{\xi} \exp[\, -S \, ]} 
      {\int {\cal D}{\xi} \exp\left[ \, -\int_0^T \xi_\mu\dot{\xi}^\mu
\, \right ]} \> = \>\exp[\, - S_{\rm cl} \, ]
\end{equation}
since the determinant from the quantum fluctuations is canceled by the
denominator. Taking the continuous limit of  $ S_{\rm cl} $ only the first 
term in Eq.~(\ref{S classic}) survives and we obtain the 
required result. Having proven the 
relation (\ref{Berezin integral}) by writing the functional integrals
in a well defined  discretized form 
we can now use it in Eq.~(\ref{G hat with diff}) with all manipulations
 formally done in the continuum.

\end{itemize}

\noindent
Using the representation
\begin{equation}
\exp \left \{\int_0^T dt \> \rho^{\mu}(t)\gamma_{\mu} \right \} \> = \> 
\exp \left \{ \gamma_{\mu}\frac{\partial}{\partial \Gamma_{\mu}}\right \}
\> \> \exp \left \{\int_0^T
dt \> \rho^{\mu}(t)\Gamma_{\mu} \right \}\Biggr|_{\Gamma_{\mu} = 0} 
\label{Weyl symbol}
\end{equation}
we obtain
\begin{equation}
U(x,y) = \exp \left ( \gamma \cdot \frac{\partial}{\partial \Gamma} 
\right ) \int{\cal D}x  \, {\cal D}p  \, {\cal D}\xi \> N^{\, \rm spin} \!
\exp \left \{ - i\int_0^T \> dt
\biggl [  p \cdot \dot x - i \xi \cdot \dot \xi + 
{\cal H}( p,x,2 \xi + \Gamma)  \biggr ]   \right \}
_{\Gamma = 0}
\label{phase space PI}
\end{equation}
Here
\begin{equation}
N^{\, \rm spin} \> = \>\left [ \> \int {\cal D}\xi \> \exp \left ( \, 
- \int_0^T dt \> \xi_{\mu} \dot \xi^{\mu} \, \right ) \> 
\right ]^{-1}
\label{N spin}
\end{equation}
is a normalization factor for the spin integral.
Note that the operation in Eq.~(\ref{Weyl symbol}) is in general 
{\it not} just a 
replacement of the boundary variable $ \Gamma $ by the corresponding 
Dirac $\gamma$ matrix but involves an antisymmetrization as well.
For example, $ \> 
\exp \left \{ \gamma \cdot \partial / \partial \Gamma \right \} 
\Gamma_{\mu} \Bigr|_{\Gamma = 0} \> = \>\gamma_{\mu} \> $, but
\begin{eqnarray}
\exp \left \{ \gamma^{\mu}\frac{\partial}{\partial \Gamma^{\mu}}\right \} 
\Gamma_{\mu} \Gamma_{\nu}\Biggr|_{\Gamma = 0} &=&\frac{1}{2}
\left ( \gamma \cdot \frac{\partial}{\partial \Gamma} \right )^2
\Gamma_{\mu} \Gamma_{\nu}\Biggr|_{\Gamma = 0} \> = \>\frac{1}{2}
\left ( \gamma \cdot \frac{\partial}{\partial \Gamma} \right ) \left ( \> 
\gamma_{\mu} \Gamma_{\nu} + \Gamma_{\mu} \gamma_{\nu} \> \right ) 
\Biggr|_{\Gamma = 0} \nonumber \\
&=& \frac{1}{2} \left ( \> - \gamma_{\nu} \gamma_{\mu} + \gamma_{\mu}
\gamma_{\nu} \> \right ) \> .
\end{eqnarray}
This is the inverse transformation of the Weyl representation for fermionic 
operators.

\end{document}